# Multi-pixel Geiger-mode avalanche photodiode and wavelength shifting fibre readout of plastic scintillator counters for the EMMA underground experiment

E. V. Akhrameev[1], L. B. Bezrukov[1], I. M. Dzaparova[1], I. Sh. Davitashvili[1], T. Enqvist[2], H. Fynbo[4], Zh. Sh. Guliev[1], L. V. Inzhechik[1], A. O. Izmaylov[1], J. Joutsenvaara[2], M. M. Khabibullin[1], A. N. Khotjantsev[1], Yu. G. Kudenko[1], P. Kuusiniemi[2], B. K. Lubsandorzhiev[1, 5*], O. V. Mineev[1], L. Olanterä[2], V. B. Petkov[1], R. V. Poleshuk[1], T. Räihä[2], B. A. M. Shaibonov[1], J. Sarkamo[2], A. T. Shaykhiev[1], W. Trzaska[3], V. I. Volchenko[1], G. V. Volchenko[1], A. F. Yanin[1], N. V. Yershov[1]

[1]*Institute for Nuclear Research of RAS, Moscow Russia*

[2]*CUPP/Pyhäsalmi, University of Oulu, Oulu Finland*

[3]*Department of Physics, University of Jyväskylä, Jyväskylä Finland*

[4]*Department of Physics and Astronomy, University of Århus, Denmark*

[5]*Kepler Center for Astro and Particle Physics, University of Tuebingen, Germany*

∗ *Corresponding author: postal address: pr-t 60th Anniversary of October, 7a, 117312 Moscow, Russia; phone: +7-095-1353161; fax: +7-095-1352268;*

E-mail: lubsand@pcbai10.inr.ruhep.ru

**Abstract**

The results of a development of a scintillator counter with wavelength shifting (WLS) fibre and a multi-pixel Geiger-mode avalanche photodiode readout are presented. The photodiode has a metal-resistor-semiconductor layered structure and operates in the limited Geiger mode. The scintillator counter has been developed for the EMMA underground cosmic ray experiment.

PACS: 29.40.Mc; 85.60.Dw; 85.60.Gz; 95.85.Ry

Key words: Plastic scintillator, cosmic rays, muon, radioactivity background, multi-pixel avalanche photodiode.

## 1. Introduction

The EMMA experiment is under construction in the Pyhäsalmi mine in central Finland [1, 2]. The aim of the experiment is to study the primary cosmic rays chemical composition at and above the "knee" region of $\sim 3 \times 10^{15}$ eV [3] by measuring underground cosmic ray muons multiplicity and their lateral distribution and arrival direction. The array will cover $\sim 150$ m$^2$ of detector area at a depth of 85 m corresponding to $\sim 240$ meters of water equivalent (m.w.e.) providing a muon energy threshold of $\sim 50$ GeV. The rock overburden array filters out all other charged particles of the air shower except high energy muons. A schematic layout of the array is shown in Fig. 1. Each of the 9 detector units has an area of $\sim 15$ m$^2$ and consists of drift chambers, previously used in the LEP-DELPHI experiment at CERN [4], and plastic scintillator counters described in this paper. The acceptance of the array, assuming an area where the shower axis can be determined with an accuracy better than 6 m, was evaluated to be approximately 300 m$^2$·sr for extensive air showers initiated by a PeV proton.

The scintillator counters of the array are a cast plastic scintillator with wavelength shifting fibre and a multi-pixel Geiger-mode avalanche photodiode readout.

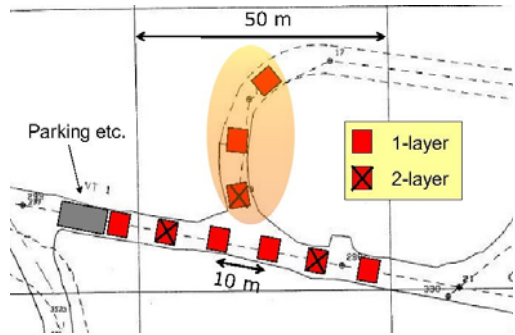

Fig. 1. The EMMA experiment layout

## 2. Multi-pixel Geiger-mode avalanche photodiode.

The multi-pixel avalanche photodiode with a metal-resistor-semiconductor layer structure operates in the limited Geiger mode. Such photodiodes known as MRS APD are produced by the CPTA Company, Moscow. The photodiodes have been selected for the EMMA experiment scintillator counter. For more details on MRS APD we refer the readers to [5-7]. The photodiode used in the scintillator counter has a 1.1 mm$^2$ sensitive area with 556 pixels of 45 × 45 μm$^2$ each grown on a common p$^+$-type silicon substrate. A sketch of the MRS APD structure is shown in Fig. 2. The main parameters of the MRS APD depend strongly on its operating voltage and ambient temperature.

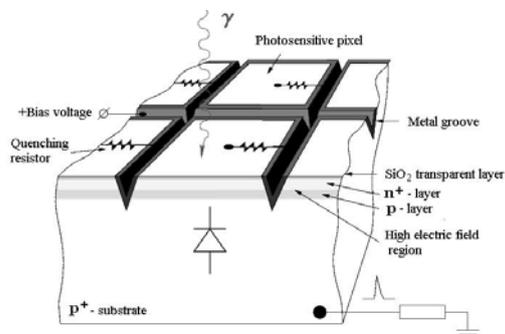

Fig. 2. A sketch of MRS APD structure

A typical operating voltage of the MRS APD selected for use in the scintillator counter is in the range of 23-35 V providing a gain of ~(2-5)×10$^5$ at room temperature. The photodiodes have an excellent single photoelectron (pe) resolution resulting in a clear separation between pe peaks in the photodiode (multi pe pulses) charge distribution - as can be seen in Fig. 3.

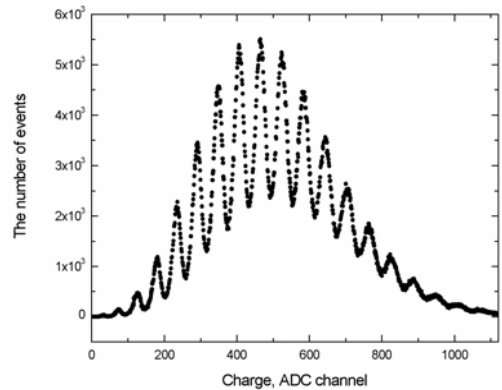

Fig. 3. Multi photoelectron charge spectrum of MRS APD

Photon detection efficiency (PDE) of the MRS APD depends on three parameters: the materials quantum efficiency, the probability for carriers to initiate Geiger discharge and a geometrical factor $\varepsilon_g$, which is the ratio between the total geometrical photodiode area to the total sensitive area of pixels. The value of $\varepsilon_g$ reaches 70% for the selected photodiodes. Measuring the photodiodes PDE one should take into account another phenomenon that effects the photodiodes sensitivity, cross-talk between pixels.

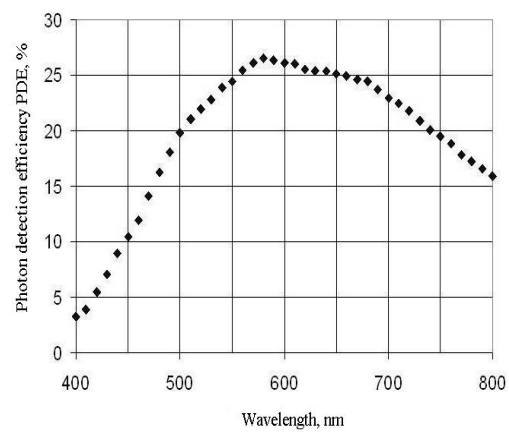

Fig. 4. Photon detection efficiency of the MRS APD versus wavelength

The cross-talk originates from the fact that some photons are produced in a pixel during the Geiger discharge development. Such photons are emitted isotropically and can penetrate into neighboring pixels engendering carriers and initiating in turn Geiger discharge there. Thus cross-talk increases the number of fired pixels in the photodiode. The cross-talk rate depends mainly on the operating voltage of the photodiode

The dependence of the photodiodes PDE on photon wavelength measured at room temperature using a spectrophotometer and a reference PMT which is well calibrated is shown in Fig. 4. The values of PDE presented in Fig. 4 are corrected for the measured values of cross-talk which is less than 10%. The photodiodes have a good sensitivity in the green region which is of utmost importance for WLS fibre readout. The photodiodes PDE is about 20-23% at $\lambda$=515 nm at room temperature and at a gain of ~5×10$^5$. The dark current counting rate of the photodiodes at this gain and temperature does not exceed 1.5-2 MHz above a threshold of ~0.5 photoelectrons (pe) and does not influence the scintillator counter performance.

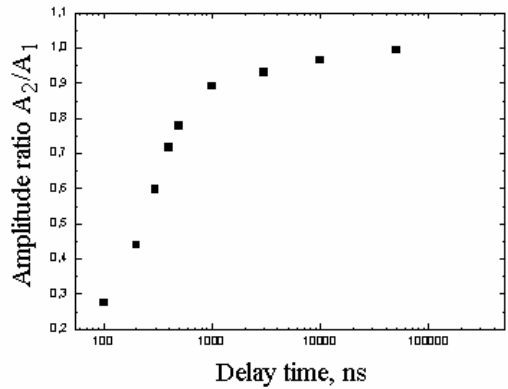

Fig. 5. Recovery time of the MRS APD

Another important parameter of the photodiode is recovery time. The recovery time of the photodiode was measured with two time-correlated light signals. The first light pulse with ~100 ps width from a laser diode (LD) with $\lambda_{max}$ = 655 nm fires all pixels of the photodiode. The second pulse from a blue LED has amplitude of ~50 pe and ~1 ns width. The delay time of the second signal was adjusted relative to the first signal. At every fixed delay time the amplitude of the photodiode responses to the LED signals with LD signals switched off and on, $A_1$ and $A_2$ respectively, were measured. The ratio between $A_2$ and $A_1$ as a function of the delay time between the signals is presented in Fig. 5. The recovery time of the photodiode is ~1µs (at the level of 0.9$A_1$), which is acceptable for the EMMA experiment where event rates will be of a few Hz per scintillator counter as shown in the next section.

The temperature in the experiment is expected to be rather stable at around 16-18°C. The temperature coefficient of the photodiode signal is ~ (-1.5÷-2)°C in the temperature range of 14-30°C. The dark current counting rate of the photodiode depends almost linearly on the temperature with a coefficient of ~60 kHz/°C.

## 3. The scintillator counters of the EMMA experiment

A photograph of a scintillator counter from the EMMA experiment is shown in Fig. 6. The scintillator counter is a polystyrene based cast plastic scintillator of 12.5×12.5×3.0 cm$^3$ size with a 3 mm deep spiral groove cut into one side. The scintillator is etched by a chemical agent resulting in the formation of a thin layer of micro pores over the scintillator surface. The thickness of the layer is of 30-100 µm depending on the etching time. A 0.5 m long Kuraray Y11 (200 ppm dopant) wavelength shifting (WLS) fiber with ~1 mm diameter core is placed in the groove and optically coupled with the scintillator bulk material by the Bicron BC 600 optical cement. One of the fiber ends is polished and connected with a MRS APD. The other end of the fibre is also polished and mirrored to increase the scintillator counter light yield. The photodiode has PDE of 23% ($\lambda$=515 nm) and gain of ~5×10$^5$ at 27.3V bias voltage and room temperature.

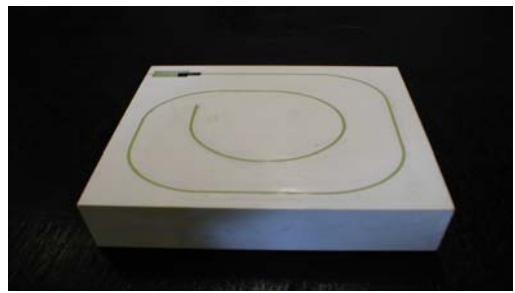

Fig. 6. Scintillator counter of the EMMA experiment.

The charge spectrum of events measured with the scintillator counter in a self-triggered mode at the surface laboratory at Pyhäsalmi at temperature of ~18°C is shown in Fig. 7. The sharp cut off in the left part of the spectrum is due to the discriminator threshold of the set-up corresponding to ~15 pe. A peak produced by cosmic ray muons can be clearly seen and corresponds to ~68 pe. The peak-to-valley ratio of the spectrum was ~2.5. The total counting rate of the scintillator counter is ~5 Hz. As illustrated in Fig. 7 the scintillator counter has a cosmic ray muon detection efficiency of nearly 100%. The total counting rate of the scintillator counter at the surface was ~7 Hz.

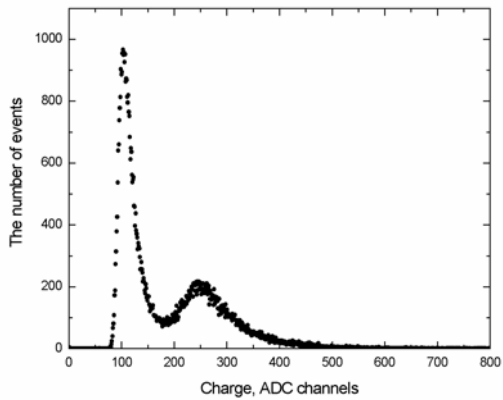

Fig. 7. Charge spectrum of events detected by the scintillator counter in the surface laboratory.

The charge spectrum of events measured with the same scintillator counter in the underground laboratory in the Pyhäsalmi mine at the depth of 85 m (240 m.w.e.) is shown in Fig. 8. The discriminator threshold was the same as in the surface measurements. The temperature in the underground laboratory has been stabilized to the level of ~18$^0$C throughout the measurements. The largest contribution to the spectrum is from background due to radioactivity in the surrounding rock. The muon peak remains at the same position as in the surface measurements corresponding to ~68 pe.

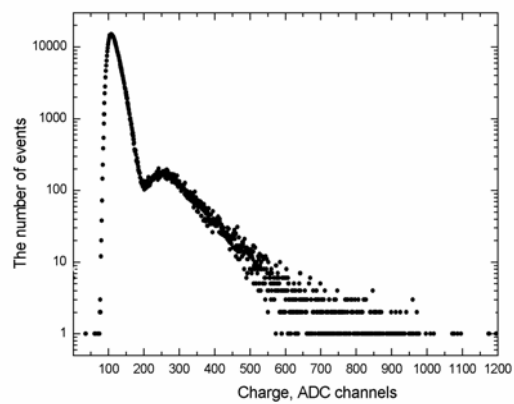

Fig. 8. Charge spectrum of events detected by the scintillator counter in the underground laboratory.

As can be noted from Fig. 8 the cosmic ray muon contribution to the total spectrum of the events detected by the scintillator counter is negligibly small at such threshold and at the same time the muon detection efficiency is sufficiently high. So, the total counting rate of the scintillator counter is ~5 Hz above the threshold of ~15 pe or 0.22 MIP and almost entirely due to radioactivity from the rock surrounding the underground laboratory.

Measurements carried out with the same scintillator counter at the Baksan Laboratory in Russia both at the surface and underground (at the same depth as in Pyhäsalmi mine) laboratories showed very similar results [8].

## 4. Conclusion

A plastic scintillator counter with WLS fibre and multi-pixel Geiger-mode avalanche photodiode readout has been developed for the EMMA underground cosmic ray experiment. The first measurements carried out with the scintillator counter in the surface and underground laboratories demonstrate its high cosmic ray muon detection and background rejection efficiencies.

## 5. Acknowledgments

The authors would like to express their gratitude to Dr. D. G. Middleton and Dr. V. Ch. Lubsandorzhieva for careful reading of the manuscript and many useful discussions and valuable remarks.

**References**

1. T. Enqvist et al. Journal of Physics: Conf. Series. 39(2006) 478.
2. T. Enqvist et al. Nucl. Phys. B (Proc. Suppl.) 175-176 (2008) 307.
3. G. V. Kulikov, G. B.Khristiansen. JETP, 1958, V.35, P.635.
4. DELPHI Collaboration, Nucl. Instr. and Meth. A 303 (1991) 233.
5. O. Mineev et al., Nucl. Instr. and Meth. A 577 (2007) 540.
6. Yu. Musienko et al., Instrum. Exp. Tech., 51 (1) (2008) 101.
7. Yu.Kudenko et al., Proceedings of Science, PoS PD07 (2009) 016.
8. V. I. Volchenko et al. arXiv: 0810.2414 [physics].